\documentclass{bmcart}

\usepackage{graphicx}
\usepackage{hyperref}
\usepackage[utf8]{inputenc} 
\usepackage{float}
\restylefloat{table}

\startlocaldefs
\endlocaldefs

\begin{document}

\begin{frontmatter}

\begin{fmbox}
\dochead{Research}


\title{Dank or Not? -- Analyzing and Predicting the Popularity of Memes on Reddit}


\author[
   addressref={aff1,aff2},                   
   email={k\_barnes@coloradocollege.edu}   
]{\inits{KB}\fnm{Kate} \snm{Barnes}}
\author[
   addressref={aff1,aff3},
   email={triesenmy@ku.edu}
]{\inits{TR}\fnm{Tiernon} \snm{Riesenmy}}
\author[
   addressref={aff1,aff4},
   email={dtrinh@haverford.edu}
]{\inits{MDT}\fnm{Minh Duc} \snm{Trinh}}
\author[
   addressref={aff1,aff5},
   email={eliriana.lleshi@tufts.edu}
]{\inits{EL}\fnm{Eli} \snm{Lleshi}}
\author[
   addressref={aff1,aff6},
   email={balogh.nora@dmlab.hu}
]{\inits{NB}\fnm{N\'ora} \snm{Balogh}}
\author[
   addressref={aff1,aff7,aff8},
   corref={aff7},
   email={molontay@math.bme.hu}
]{\inits{RM}\fnm{Roland} \snm{Molontay}}


\address[id=aff1]{
  \orgname{Aquincum Institute of Technology}, 
  \city{Budapest},                              
  \cny{Hungary}                                    
}
\address[id=aff2]{%
  \orgname{Colorado College},
  \city{Colorado Springs},
  \cny{USA}
}

\address[id=aff3]{%
  \orgname{University of Kansas},
  \city{Lawrence},
  \cny{USA}
}

\address[id=aff4]{%
  \orgname{Haverford College},
  \city{Haverford},
  \cny{USA}
}

\address[id=aff5]{%
  \orgname{Tufts University},
  \city{Medford},
  \cny{USA}
}

\address[id=aff6]{%
  \orgname{Dmlab Ltd.},
  \city{Budapest},
  \cny{Hungary}
}

\address[id=aff7]{%
  \orgname{MTA-BME Stochastics Research Group},
  \city{Budapest},
  \cny{Hungary}
}

\address[id=aff8]{%
  \orgname{Department of Stochastics, Budapest University of Technology and Economics},
  \city{Budapest},
  \cny{Hungary}
}



\end{fmbox}


\begin{abstractbox}

\begin{abstract} 
Internet memes have become an increasingly pervasive form of contemporary social communication that attracted a lot of research interest recently.
In this paper, we analyze the data of 129,326 memes collected from Reddit in the middle of March, 2020, when the most serious coronavirus restrictions were being introduced around the world. This article not only provides a looking glass into the thoughts of Internet users during the COVID-19 pandemic but we also perform a content-based predictive analysis of what makes a meme go viral. Using machine learning methods, we also study what incremental predictive power image related attributes have over textual attributes on meme popularity. We find that the success of a meme can be predicted based on its content alone moderately well, our best performing machine learning model predicts viral memes with AUC=$0.68$. We also find that both image related and textual attributes have significant incremental predictive power over each other. 
\end{abstract}


\begin{keyword}
\kwd{memes}
\kwd{popularity prediction}
\kwd{machine learning}
\kwd{sentiment analysis}
\kwd{image analysis}
\kwd{content-based analysis}
\kwd{social media}
\kwd{visual humor}
\kwd{covid-19}
\end{keyword}


\end{abstractbox}
%

\end{frontmatter}



\section*{Introduction}
Over the past decade, Internet memes have become a pervasive phenomenon in contemporary Web culture~\cite{laineste2017laughing}. Due to their popularity, memes have received considerable attention in areas such as pop culture, marketing, sociology, and computer science~\cite{bauckhage2013mathematical,journell2019political}. In the time of the COVID-19 pandemic, memes have become an even more important part of social life since due to social distancing orders more people turned to the Internet for everyday interactions. As a result, Web culture is moving faster than ever and social media sites have exploded with coronavirus memes as people all over the world try to take this serious situation with a pinch of humor~\cite{bischetti2020funny}.

The increasingly participatory nature of the Internet has made memes into a social phenomenon, created, altered, and spread by Internet users themselves. Today, memes are not only a source of humor but also draw attention to poignant cultural and political themes \cite{brodie2009memebook}. Memes tend to reflect pressing global issues and while they are not always loyal to the facts \cite{simmons2011quotes}, they often show what the public is noticing most. Many authors have explored the social network factors that lead a meme to go viral but bracketed the impact that meme content may have on popularity  \cite{gleeson2015popularity,gleeson2014criticality,weng2012competition}. In other areas of human achievement viral success is closely linked with merit \cite{barabasi2016success}, but it is unclear what characteristics lead a meme to have merit. This paper investigates the relationship between a meme's content, excluding social network features, and its popularity. Along the way, it exposes what topics were popular on the Internet during the global COVID-19 pandemic.

Our paper joins a growing body of literature that employs network science and data science techniques to predict the popularity of Internet memes  \cite{weng2012competition,maji2018net,tsur2015don,wang2011epidemiological}. Here we analyze the popularity of coronavirus memes based on 129,326 records scraped from Reddit, the largest social news and entertainment site.
The main contributions of this work can be summarized as follows:
\begin{itemize}
    \item Using advanced machine learning techniques (such as convolutional neural networks, gradient boosting, and random forest), we perform a content-based analysis of what makes a meme successful, considering several features from both text and image data.
    \item We stand apart from other authors by investigating whether the success of a meme can be predicted based on its content alone, excluding social network factors.
    \item We not only study what makes a meme viral, but we also analyze what incremental predictive power image related attributes have over textual attributes on memes popularity.
    \item Our study provides a looking glass into the thoughts of Internet users during the COVID-19 pandemic.
\end{itemize}

\section*{Related work}

The term ``meme" precedes the digital age, stemming from the Greek mimēma, something imitated. Thus, memes are pieces of cultural information that remain relatively unchanged as they are passed between individuals in society through imitation. In the modern age, the term has been co-opted by Internet users to mean snippets of information that self-replicate on the Internet \cite{dawkins2016selfish,shifman2014memes}. When memes took the form of hashtags, tweets, photos, quotes or jokes shared repetitively on the web they became highly visible and a common source of data for social computer science researchers. They are transmitted from person to person through social media sites, online news, or blog posts and can reach extremely large audiences in short amounts of time. These viral memes are important, shared social phenomena. They can represent common opinions, cultural norms \cite{dynel2020swiss}, carry political power or motivate social change \cite{dynel2020incel,simmons2011quotes,mcclure2016politics,du2020policontent}. Humorous content may play a crucial role in the spread of memes as it encourages user interaction and creates a sense of in-group connection  \cite{vasquez2019humourbook}. However, little is understood about what kind of information is so appealing to Internet users as to become \textit{viral}. Ours is among few studies that places the content of memes under scholarly analysis.

The journey of Internet popularity is commonly framed in network science as competition between memes for limited user attention \cite{gleeson2015popularity, gleeson2014criticality}. Memes are analogous to genes \cite{wang2011epidemiological}, cultural fragments passed down through generations. In itself, the Darwinian frame through which memes are understood recognizes the importance of meme content. However, most studies focus on how memes diffuse through online social networks \cite{wang2011epidemiological} taking into account user interests \cite{weng2012competition}, memory \cite{gleeson2015popularity}, and other social factors.  

Many studies have successfully predicted the viral Internet memes based on social network factors \cite{maji2018net,weng2014net} and others have designed mathematical models that closely align with the actual transmission of memes through the Internet \cite{weng2012competition,wang2011epidemiological,bauckhage2011insights}. Even when measured in many ways, meme popularity displays a long-tailed distribution. Few memes actually become viral, and most are only appreciated by a few tens of people \cite{gleeson2015popularity}. Memes distributed in more diverse and well-connected audiences are more likely to go viral~\cite{weng2014net}. Additionally, people are more likely to share memes related to content that they have shared in the past \cite{weng2012competition}. All of these studies put forth neutral models: they assumed no inherent advantage in terms of memes' attractiveness to individuals.

In addition to social network factors, the content and formatting of a meme can effect its popularity. Tsur \textit{et al.} analyzed hashtags on Twitter, and found that brevity is the most important feature for the memes' popularity followed by certain legibility characteristics such as capitalization \cite{tsur2015don}. Berger and Milkman found that more emotionally arousing text segments from online news are more likely to go viral~\cite{berger2012arousing}. Our analysis of meme captions' length replicates the finding by Tsur \textit{et al.} but our meme sentiment analysis differs from the Berger and Milkman finding. Others note that there is ever increasing engagement with political memes among adults on the Internet and express concern that political memes will be used to promote extremism or spread misinformation. According to a recent look at Twitter data, 30 percent of image-with-text memes contain political content \cite{du2020policontent}. There are also disparities among what political and demographic groups share those memes ~\cite{mcclure2016politics}. 

While many papers investigate short text data like hashtags \cite{tsur2015don,weng2014net}, quotes \cite{simmons2011quotes}, and Google searches \cite{wang2011epidemiological}, few look at the combination image-with-text memes we consider here. Qualitative studies describe the symbols used in meme sub-genres and how their used but do not analyze the impact of these symbols on the memes popularity \cite{dynel2016animal,dynel2020swiss}. A study by Bauckhage \textit{et. al.} models how users' attention to image-with-text memes fluctuates over time~\cite{bauckhage2013mathematical}. It shows that evolving memes (slightly different versions of the same meme) are more likely to gain popularity and stay popular for longer. Du \textit{et al.} only study the text within image-with-text memes, claiming that the image is merely a neutral background or further emphasizes information already addressed in the text \cite{du2020policontent}. Our paper contests this claim. Another study by Khosla \textit{et al.} investigates the content and social contexts of popular images alone, using data from Flickr \cite{khosla2014image}. They found that certain colors, low-level image properties like hue, and represented objects correlate with increased image popularity. However, popularity on a photo-appreciation sight like Flickr is much different than the social undertones that go into memes. In our model, similar features to those considered by Khosla \textit{et al.} show different relationships to image popularity.

Our study considers the widest array of content-based attributes in image-with-text memes so far. Furthermore, our data represents the intense political moment at the start of the global coronavirus pandemic.

\section*{Data description and preparation}
\label{sec:data}

All data for this project were collected from Reddit, the so called “front page of the Internet.” More precisely, the image-with-text memes came from the largest meme subreddits, namely \textit{r$/$MemeEconomy},  \textit{r$/$memes},  \textit{r$/$me\_irl}, \textit{r$/$dankmeme}, and  \textit{r$/$dank\_meme}. The subreddits represent communities devoted to the creation of memes and consequently, the development of a shared sense of humor on Reddit. Most popular Internet content first went viral on Reddit, hence the websites catchphrase, so popular memes from these subreddits are likely representative of the content on many other Internet sites too. Additionally the strict etiquette implemented by the Reddit community and moderators ensures that posts align with the subreddit description~\cite{sanderson2013reddit}. Thus, only image-with-text memes populate the five subreddits from which we scraped data.    

We employed the Pushshift API \cite{baumgartner2020pushshift} to scrape data from posts in the five meme subreddits. In total we scraped 129,326 unique posts from March 17th, 2020 to March 23rd, 2020 which constituted the beginning of the global coronavirus outbreak. For each post we retrieved the features found in lines 1 to 10 of Table~\ref{tab:features_table}. Additional features such as urls to access the post on Reddit and unique meme ids were scraped as well, but only the features we use for analysis are included in Table \ref{tab:features_table}. Likewise, the features downvotes, meme awards, and posting author were scraped from Reddit and eliminated early on because they were incomplete, populated mostly with zeros. Many of the features scraped from Reddit metadata were already numerical, such as created\_utc and ups. The categorical features is\_nsfw and subreddit were one-hot-encoded into a numerical representation. 

We further processed the meme images, titles, and text from the images to enrich our feature set with more content-based features. These extracted features are listed in lines 13 to 22 of Table~\ref{tab:features_table} and discussed in more detail in the \nameref{sec_model} section. In the process of extracting the content-based features, we made a GET request on each link and observed the status code. Any post with a link that returned a 404 or other similar error was removed from the data set in order to avoid evaluating dead links. Further, any post with a media other than images, such as gifs, was removed as we only wished to consider image-based memes. These cleaning steps resulted in a total of 80,362 records for training and testing the machine learning models. After numerically encoding the image and text based content features that will be discussed in detail in the next couple sections, there were a total of 97 data attributes.

\begin{table}[H]
\centering
\begin{tabular}{|l|l|l|l|}
\hline
        &   \textbf{Feature}                    & \textbf{Type}        & \textbf{Description}                   \\ \hline
\textbf{1} & \textbf{created\_utc}        & utc timestamp        & time of post submission                \\ \hline
\textbf{2} & \textbf{ups}                 & integer               & number of upvotes received     \\ \hline
\textbf{3} & \textbf{is\_nsfw}            & boolean              & indicates if only suitable for 18+     \\ \hline
\textbf{4} & \textbf{subreddit}           & string               & subreddit of the submission            \\ \hline
\textbf{5} & \textbf{subscribers}         & integer              & number of subscribers to the subreddit \\ \hline
\textbf{6} & \textbf{thumbnail.height}    & floating point value & height of the thumbnail                \\ \hline
\textbf{7} & \textbf{thumbnail.thumbnail} & string               & thumbnail media                        \\ \hline
\textbf{8} & \textbf{thumbnail.widith}    & floating point value & width of thumbnail                     \\ \hline
\textbf{9} & \textbf{title}               & string               & title of the submission                \\ \hline
\textbf{10} & \textbf{media}               & string               & link to associated meme media                \\ \hline
\textbf{11} & \textbf{ups\_normed}     & floating point value               & ups normalized with subscribers       \\ \hline
\textbf{12} & \textbf{dankornot}         & integer      & label ups\_normed for binary classification \\ \hline
\textbf{13} & \textbf{processed\_words}    & list of strings      & filtered and stemmed words from title and image \\ \hline
\textbf{14} & \textbf{word\_count}         & integer      & number of words in title and image \\ \hline
\textbf{15} & \textbf{TextLength}         & integer      & number of characters in title \\ \hline
\textbf{16} & \textbf{Sentiment}         & floating point value      & text valence score \\ \hline
\textbf{17} & \textbf{avg\_H}         & floating point value      & average HSV hue value of meme \\ \hline
\textbf{18} & \textbf{avg\_S}         & floating point value      & average HSV saturation value of meme \\ \hline
\textbf{19} & \textbf{avg\_V}         & floating point value      & average HSV value value of meme \\ \hline
\textbf{20} & \textbf{30 colors}         & floating point value      & normalized pixels of color in image \\ \hline
\textbf{21} & \textbf{VGG\_features}         & list of strings      & VGG-16's  first three guesses about image content \\ \hline
\textbf{22} & \textbf{VGG\_probs}  & list of floating point values  & the probabilities of the VGG-16's first three guesses \\ \hline
\end{tabular}
\caption{Features extracted from Pushshift API together with processed features}
\label{tab:features_table}
\end{table}

\begin{figure}[h!]
\centering
\includegraphics[width=0.95\textwidth]{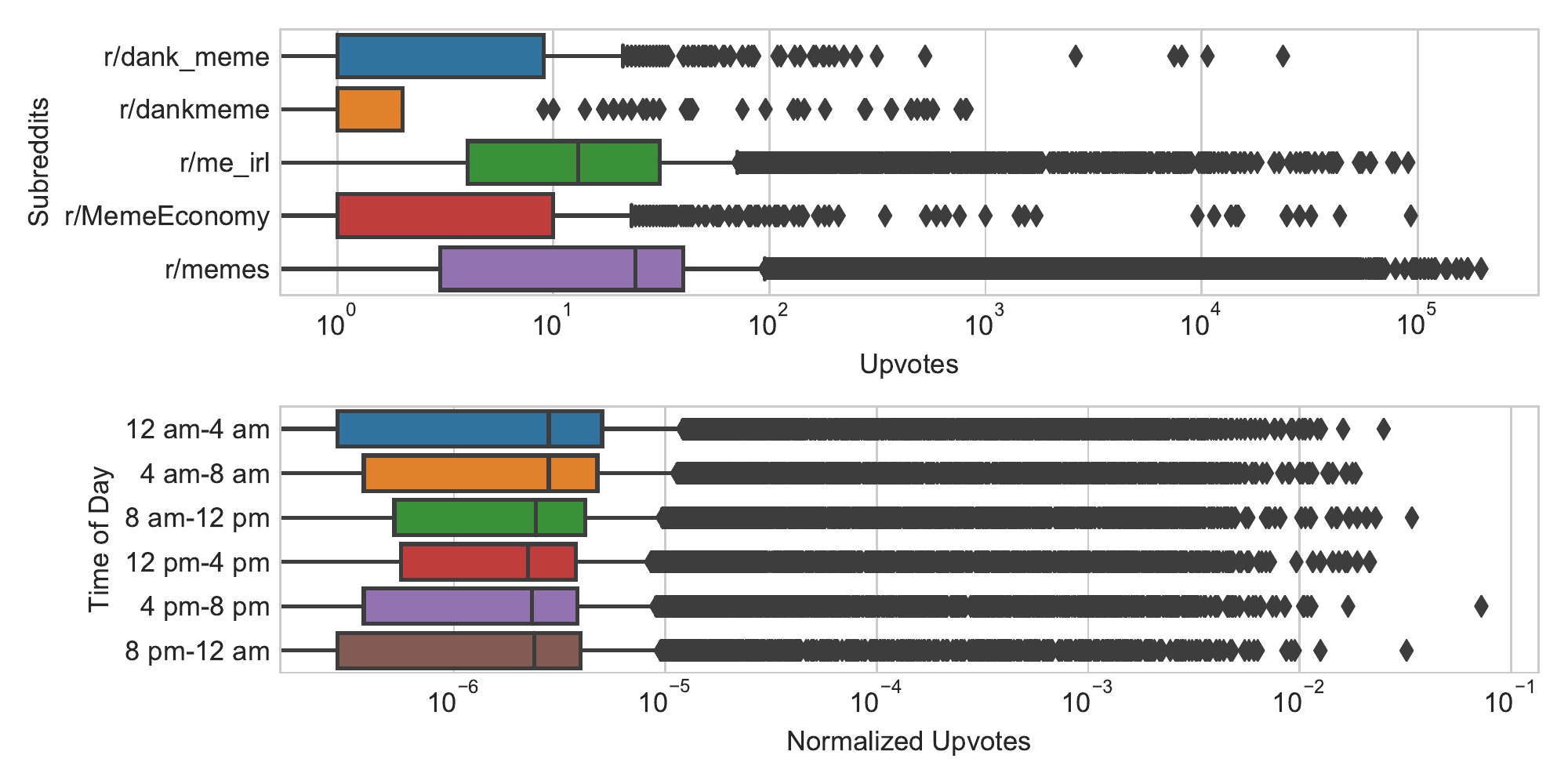}
\caption{Upper figure shows the number of upvotes for each scraped subreddit. Bottom figure presents the normalized upvotes for each time frame.}
\label{fig:reddit metadata}
\end{figure}

Some conclusions could be made based on the Reddit metadata alone. The \textit{created\_utc} feature contains the timestamp when the post appeared on Reddit in the Coordinated Universal Time zone (UTC). Since most active Reddit users reside in the USA \cite{statistica2020reddit}, we converted this to  North American Central Time Zone. Based on this feature we created a categorical feature representing the time of day, in four hour increments, when the post was created. The bottom subfigure of Figure~\ref{fig:reddit metadata} shows the effect of time of day on the normalized upvotes that posts received. Posts published on Reddit from midnight to noon Central US time have a higher chance to attract great attention. This result could mean that most upvotes on Reddit are accumulated during the course of the day, in USA time zones. The memes posted during daytime (Central US time) have more chance to receive moderate attention, while memes posted at night are more exposed to extreme events, meaning that receiving very low attention or great attention. The observation that memes posted at night have more chance to be dank is in line with the phenomenon that was observed by Sabate \textit{et al.} based on the analysis of popularity of Facebook content~\cite{sabate2014factors}. The authors argue that if content is posted during periods with low user activity (at night), when users will connect in peak hours the post appears at the top of the news wall, that makes it more likely to be liked, commented or shared.

The more subscribers, the more social exposure, so the number of upvotes a post received was likely influenced by the number of subscribers to the subreddit where it was posted. In our data, \textit{r$/$memes} has the most subscribers, around 10,000,000, followed by \textit{r$/$me\_irl} with around 4,000,000, and  \textit{r$/$Meme\_Economy} with around 1,000,000 subscribers; \textit{r$/$dank\_meme} and \textit{r$/$dankmeme} have the least subscribers, less than 500,000 and less than 1,000 subscribers respectively. Indeed, we can observe a positive correlation between upvotes and subscribers, as the subreddits with more subscribers tend to get more upvotes (see Figure \ref{fig:reddit metadata} upper subfigure). To confirm this observation, we determined the median number of upvotes for each respective subreddit and calculated a Pearson correlation coefficient between these values and the number of subscribers for each subreddit.  We received a value of 0.977 for the Pearson correlation coefficient, which indicates a strong near-linear relationship between the upvotes of a post and the number of subscribers to the subreddit. To eliminate this network effect, we normalized the number of upvotes by dividing by the number of subscribers from the respective subreddit where it was posted. In modifying the upvotes feature, we were able to better gauge the popularity of a meme based upon its content alone. 

\begin{figure}[h!]
\centering
\includegraphics[width=0.95\textwidth]{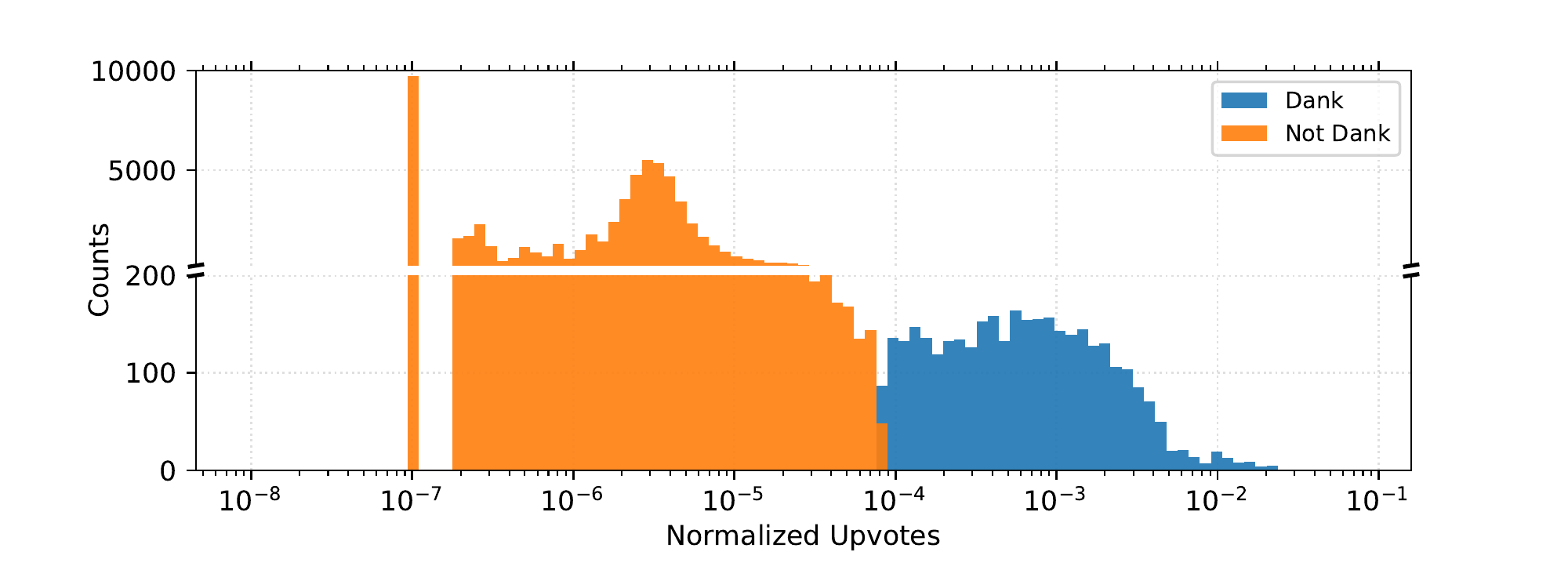}
\caption{The distribution of the normalized upvotes for dank and not dank memes.}
\label{fig:normed_ups_distribution}
\end{figure}

The viral nature of image-and-text memes on Reddit makes this data well suited for a binary classification task. The distribution of normalized upvotes follows a long-tailed distribution:  most memes received few upvotes while few memes received many upvotes as shown in Figure \ref{fig:normed_ups_distribution}. Therefore, viral memes usually differ by two or more orders of magnitude from not viral memes, as defined by our binary classification label, called \textit{dank or not} in Table \ref{tab:features_table}, and used for the supervised learning models. Using the the normalized upvotes feature as our criteria, any posts with a normalized upvotes value in the top 5\% of all posts was classified as \textit{dank} (positive label, 1), and the rest were classified as \textit{not\_dank} (negative label, 0). Our data set contains 4019 \textit{dank} entries, and 76343 \textit{not\_dank} entries. Formulating our prediction labels in this way assured that we investigate the phenomenon of viral popularity (rather than moderately successful or mediocre memes) as proposed in the introduction.

We will use three supervised learning models to predict whether memes fall into the dank or not dank categories: Gradient Boosting, Random Forest, and Convolutional Neural Network models. The former two use the entire feature set described in Table~\ref{tab:features_table} for training (except the media link feature). The neural network model uses only the meme images, accessed via the media feature, as its input and it is based on a smaller sample of data records. This subset of data will be discussed in more detail in the \nameref{transfer} section.

\section*{Models and results}
\label{sec_model}

In this section we present the results of our analysis. First, an explanatory analysis is provided for the textual and image related attributes with a focus on the impact they have on meme popularity. We also present feature engineering steps. Next, we briefly describe the applied machine learning models together with their performance in predicting the success of memes. 

\subsection*{Text analysis}
\label{sec_text}

A large portion of the humor and meaning of memes are contained in the text which appears inside a meme image. This text differs from the caption of the meme which was written by the user who created the post and can be scraped directly from Reddit. Both the caption and the text contained within the meme itself may affect popularity. In this section, we study the predictive power of the attributes derived from the caption and the text extracted from the images on meme popularity.

\begin{figure}[h!]
\centering
\includegraphics[width=0.95\textwidth]{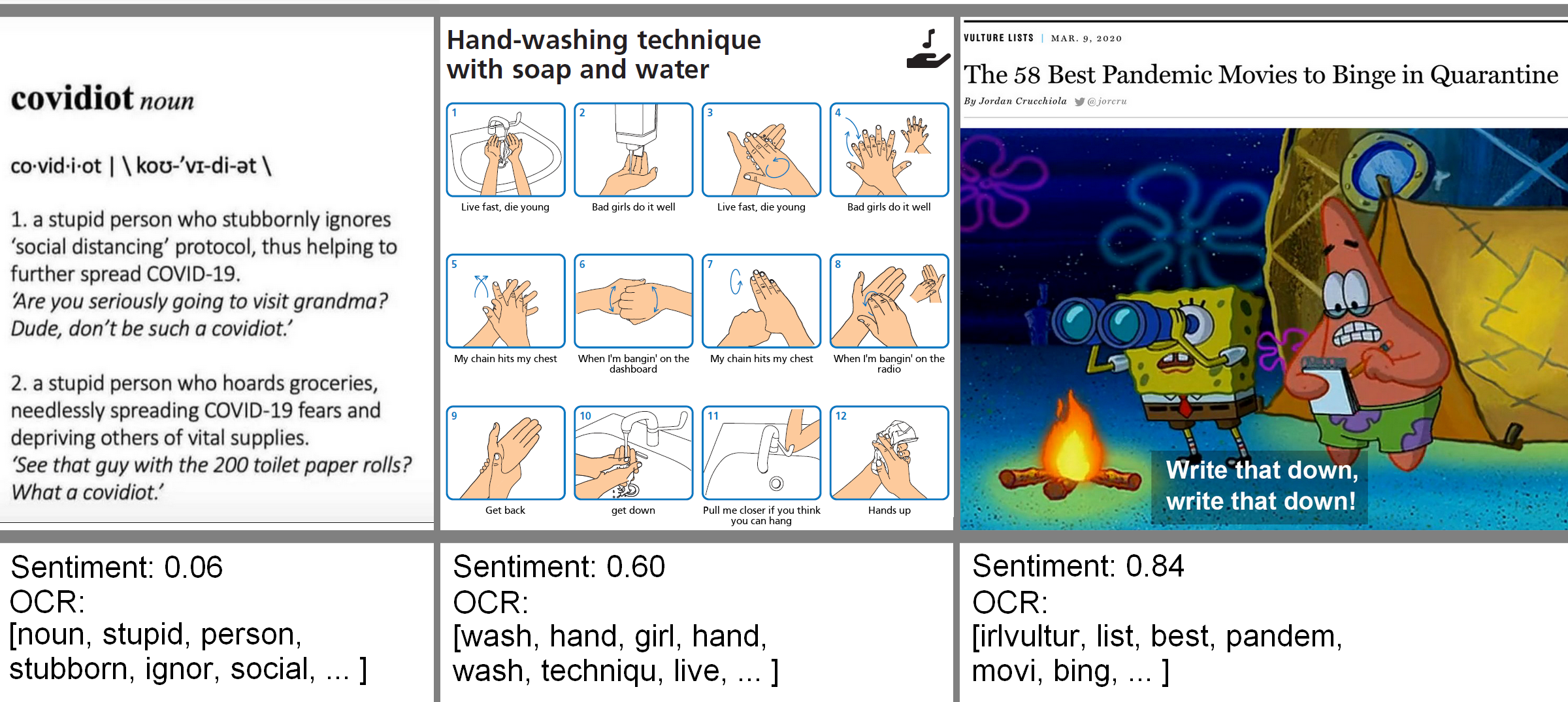}
\caption{Examples for negative, neutral, and positive sentiment memes along with a few example words obtained from OCR.}
\label{fig:sentiment_and_ocr}
\end{figure}

The text from the images was extracted using Optical Character Recognition (OCR) \cite{OCR_API}. We combined the text obtained by OCR with the caption of the meme, to gather all text associated with a meme.  Then we performed tokenization, lemmatization, and stemming to simplify all of the words. This was done using the NLTK and gensim Python libraries \cite{gensim2011,loper2002nltk}. Tokenization is used to split the text into a list of words, make all characters lowercase, and remove punctuation. Words that have fewer than 3 characters and stopwords were removed. Words were lemmatized so that all verbs occur in their first person, present tense form. Finally, words were stemmed, or reduced to their root form. For example, the \textit{``processed words"} extracted from three memes using  OCR, tokenization, lemmatization, and stemming can be seen in Figure \ref{fig:sentiment_and_ocr}.

Using the processed text data we can extract some potentially predictive attributes such as sentiment and word count. First, we calculated the sentiment scores that quantify the feeling or tone of the text \cite{liu2012sentiment}. If the text is positive or happy, it scores closer to 1, and negative or sad texts score closer to 0. Examples for different sentiment scores are shown in Figure \ref{fig:sentiment_and_ocr}. The sentiment model we used to analyze the processed meme text uses a recurrent neural network known as LSTM (Long Short Term Memory) \cite{sentimentGithub}. This network remembers the sequences of past words in order to make predictions about the sentiment of new words. The model was trained on dictionaries with hundreds of thousands of words that were already scored for sentiment. 

\begin{figure}[h!]
\centering
\includegraphics[width=\textwidth]{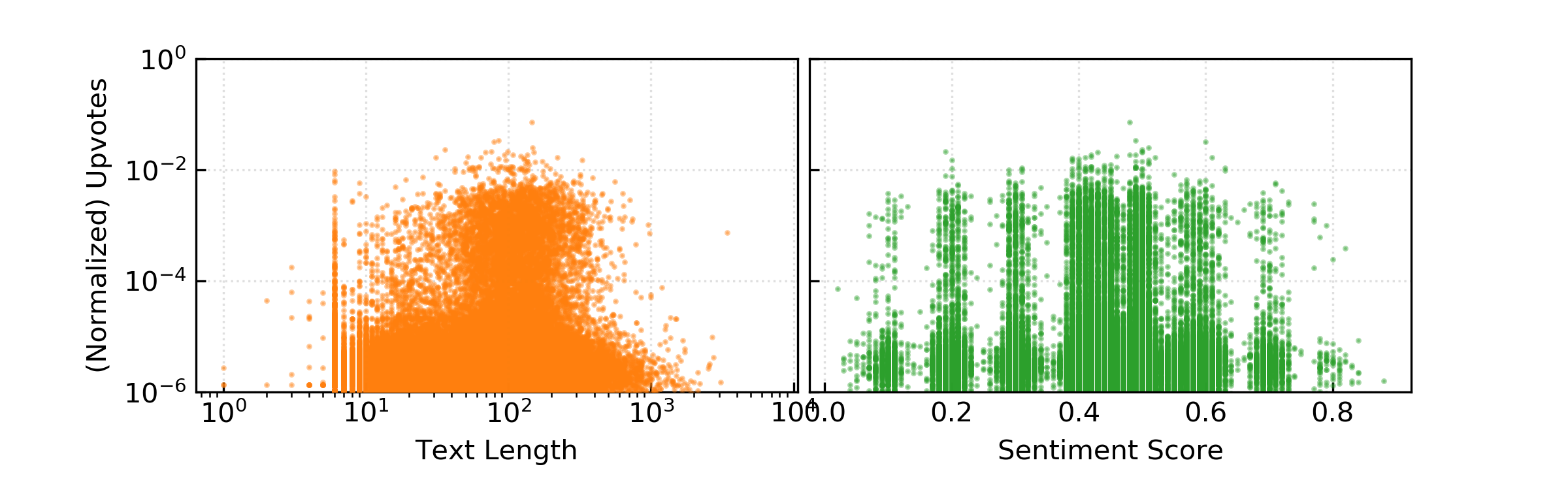}
\caption{The relationship between the extracted text features and the normalized upvotes.}
\label{fig:plots}
\end{figure}

Figure \ref{fig:plots} illustrate the relationship between the extracted text features and the normalized upvotes. The framework in which memes compete for limited user attention suggests that users may respond best to memes with shorter texts. Indeed, we found that the amount of text a viewer is required to read correlates negatively with upvotes. This is in alignment with the findings of de Vries \textit{et al.} on the popularity of brand posts in the social media \cite{de2012popularity}.

In Figure \ref{fig:plots} we can also observe that neutral memes perform better than extreme ones, but of the extremes, negative sentiments perform better than positive sentiments. These results contradict previous finding that online news content which evokes high arousal, especially negative, is more viral than neutral content \cite{berger2012arousing}. Another paper found that popular memes tend to be unique, in terms of sentiment and other features, whereas memes that are similar to most other memes perform poorly~\cite{coscia2014similarity}. It is unlikely that humor is usually helped by neutral charged content, instead its associated with surprise which is related to arousal \cite{chandrasekaran2015humor}. This result suggests that the jokes in memes particularly are about mundane, not arousing topics.

The words extracted from the text were encoded as numerical attributes and analysed for their relationship with meme content and popularity. Similar groups of words such as ``coronavirus", ``virus", and ``pandemic" were grouped together under one name. The 7 word categories can be viewed in Table \ref{tab:word_categories}. Then, these categories, along with the top 28 most frequently occurring words in the \textit{processed\_words} attribute (in Table \ref{tab:features_table}), were one hot encoded into 35 numerical feature attributes. In total, including text length, word count, and sentiment scores, there are 38 numerical text features. 

\begin{figure}[h!]
\centering
\includegraphics[width=0.9\textwidth]{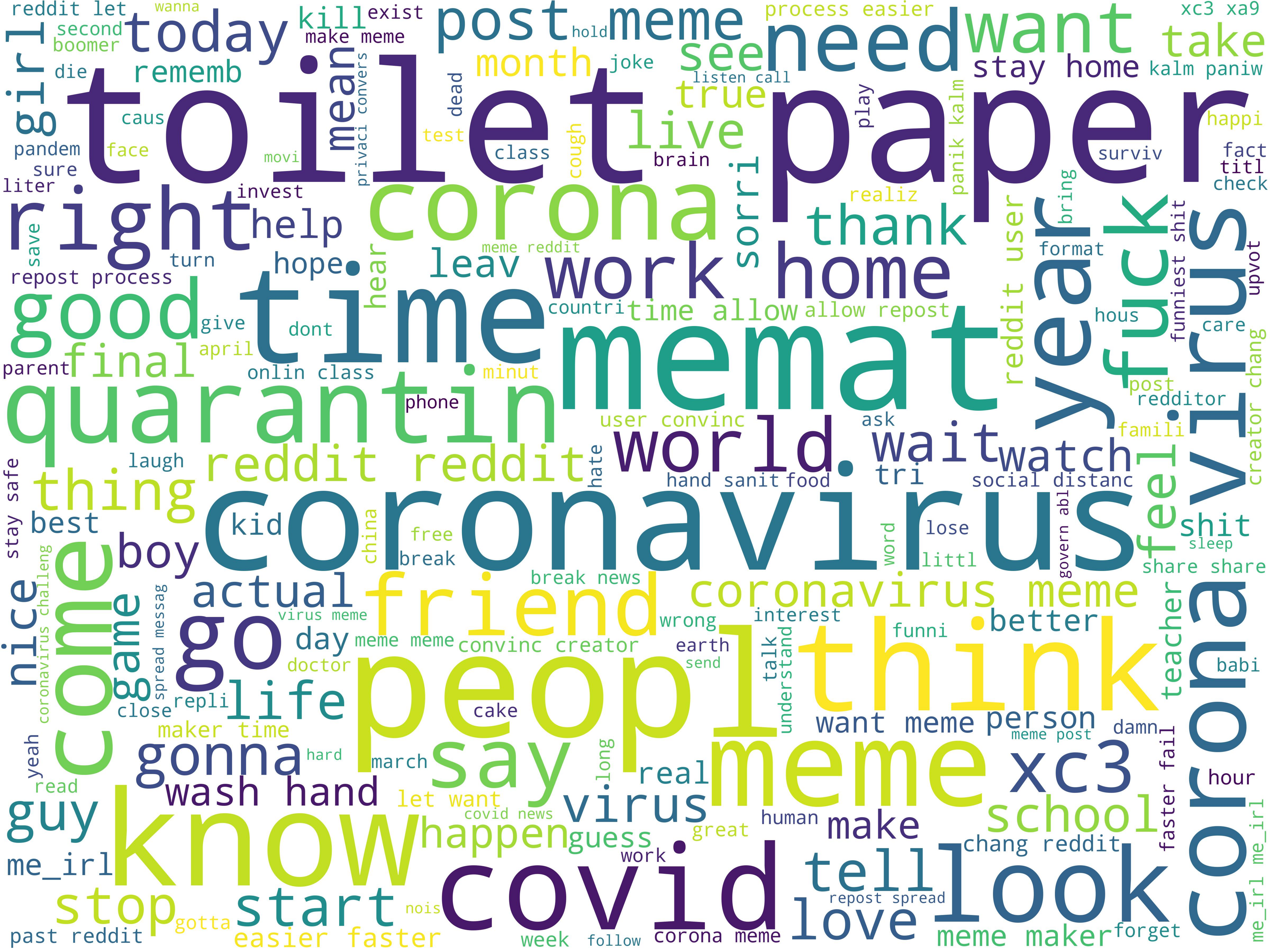}
\caption{\csentence{Wordcloud generated from all text in our scraped memes}}
\label{fig:wordcloud}
\end{figure}

A word cloud in Figure \ref{fig:wordcloud} created from every word we gathered indicates certain topics are especially prevalent in the memes from late March, 2020. For instance, ``coronavirus", ``toilet paper", ``quarantine", ``work", ``home", ``school", and ``friend" all appear most prominently in the word cloud, though some appear in slightly different versions due to our processing.  ``Memat" is notably one of the prominent words in the word cloud.  A popular meme-making website entitled mematic is used by many Reddit users, and each meme produced from the website contains a mematic watermark to indicate its origin.  The watermark was apparently read by the OCR as text from the meme. Hence, ``memat" is one of the more prominent texts found among the memes. The largest words in Figure \ref{fig:wordcloud} shows that current events do play a great role in the content of memes, though whether this sort of content has a great effect on popularity is another question. An initial analysis showed than in most cases, the words included in table \ref{tab:word_categories} are just as prevalent in the top 5\% viral memes as in non-viral memes. The largest difference we found was for the category COVID-19 synonyms in which 23\% of dank memes contained at least one word from the category and 17\% of not dank memes contained a word from that category. We aim to answer this question further in the following sections by studying the importance of these features to machine learning models.

\begin{table}[]
\begin{tabular}{|l|l|}
\hline
Category name      & Content                                                                                                                                                                                                                                                                      \\ \hline
current politics   & \begin{tabular}[c]{@{}l@{}}'econom', 'world','global','emperor','countri', 'trump', 'crash','berni',\\ 'dollar', 'stock', 'profit','market', 'bailout', 'sander', 'senat', 'democrat', \\ 'presidenti', 'debat','govern','congress', 'pass', 'govern','privaci'\end{tabular} \\ \hline
temporal moment    & '2020','time','year', 'month', 'week', 'day'                                                                                                                                                                                                                                 \\ \hline
covid culture      & 'distanc', 'social', 'quarantin','isol','hand', 'sanit','tp', 'toilet', 'paper'                                                                                                                                                                                              \\ \hline
synonyms, sick     & \begin{tabular}[c]{@{}l@{}}'fever', 'cough', 'short','sick', 'health','outbreak','exposur','breath', \\ 'diseas', 'transmiss', 'symptom', 'ill', 'infect','cough'\end{tabular}                                                                                               \\ \hline
synonyms, covid-19 & 'corona', 'coronaviru', 'viru', 'vaccin','covid-19', 'covid','outbreak', 'pandem'                                                                                                                                                                                            \\ \hline
pronouns           & \begin{tabular}[c]{@{}l@{}}'we', 'us', 'our', "we'r", 'i', "i'm", 'i’m', 'my', "i'll",'you', "you'r", 'you’r', \\ 'your', 'u', 'y’all'\end{tabular}                                                                                                                          \\ \hline
about memes        & 'meme', 'reddit', 'repost', 'comment', 'upvot', 'redditor', 'post'                                                                                                                                                                                         \\ \hline
\end{tabular}
\caption{Description of word category features}
\label{tab:word_categories}
\end{table}

\subsection*{Image analysis}
\label{sec_image}
Most images on the Internet are not neutrally charged. Subtle differences in color, definition or setting can convey vastly different meanings to the viewer. In general unique, bright, high definition images with a low depth of field are ranked more aesthetic by viewers \cite{datta2006aesth}. Additionally, the presence of certain objects in a photo lead to greater or lesser popularity on Flickr \cite{khosla2014image}. However, memes often have comedic, relatable or reactionary value which is not necessarily aesthetic. The importance of image features may differ for memes as opposed to other types of Internet images.
    
The image is an important part of a meme. An initial analysis of thumbnail area in our data showed that the majority of memes had the largest thumbnail size available on Reddit. The more popular memes also tend to have larger thumbnail areas. 

In addition to thumbnail area, we looked at the colors present in the most popular meme images. Color and thumbnail area are examples of simple image features, aspects of an image that are easily interpreted by human viewers. The colors that the human vision system perceives as distinct have larger value coordinates in HSV (hue, saturation, value) color space, therefore we extracted colors from the HSV versions of our meme images. We used an OpenCV image segmentation technique \cite{stone2018python} to isolate 30 colors, including a small range around the specific HSV value of the color. This range was used to mask the images, revealing only pixels within that color range. The number of pixels in the mask was normalized for images of differing sizes by dividing by the total number of pixels in the image. These color attributes represent the amount of each of the 30 given colors present in the meme images.

\begin{figure}[h!]
\centering
\includegraphics[width=\textwidth]{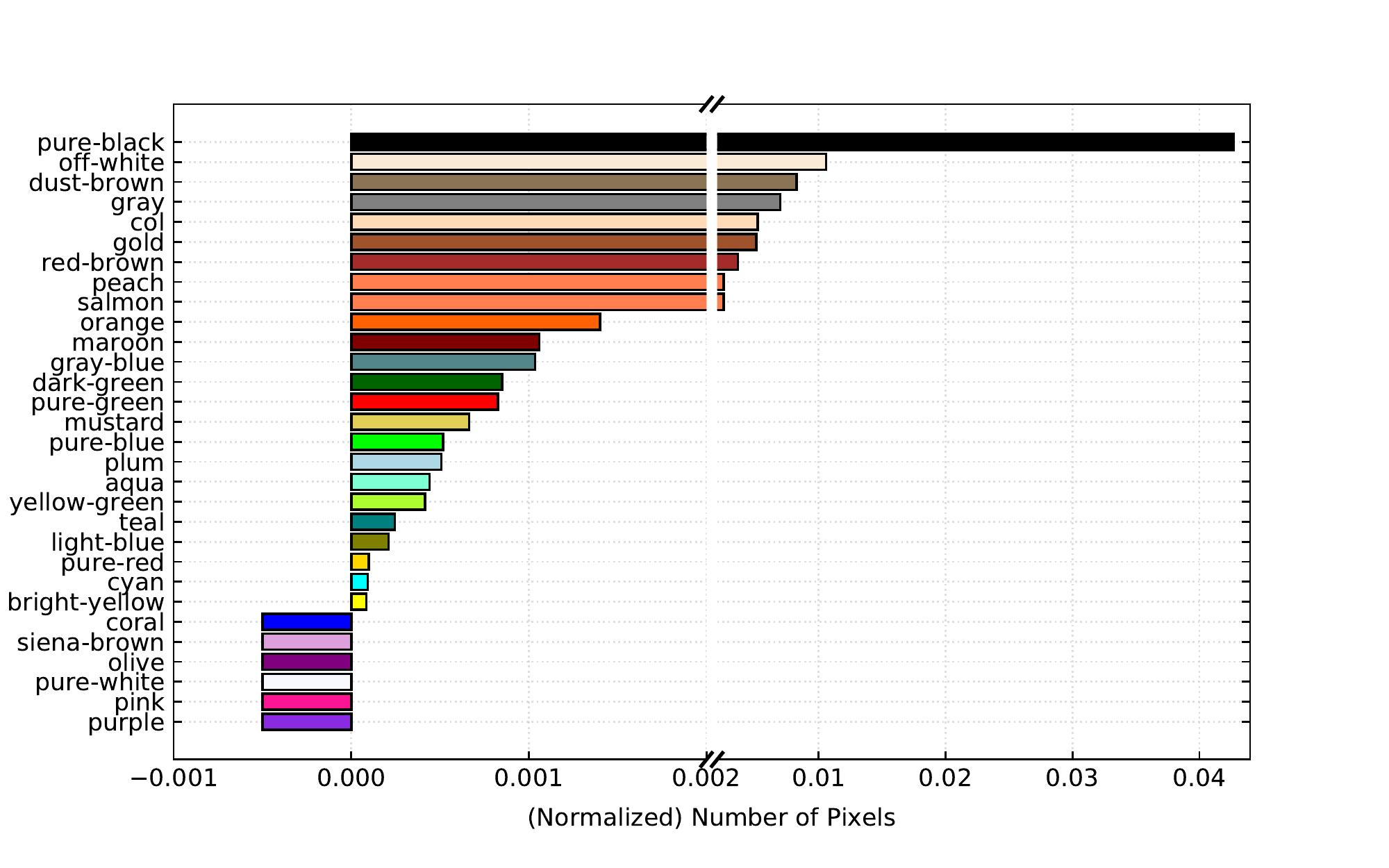}
\caption{\csentence{Color content of popular memes.}
      The average amount of each color attribute in the top 5 percent of memes, 3728 records, with the most normalized upvotes. Bars that go below 0 indicate that none of that color was present in the dank memes. }
\label{fig:dank_colors}
\end{figure}

Figure \ref{fig:dank_colors} shows the amount of each color attribute in the upper 95 percentile of popular memes. In general, muted colors are more abundant than bright colors in viral memes. Perhaps because memes tend to be mundane photos, often blurry in self-made way, unlike professional photography. This result differs from the similar analysis done by Khosla \textit{et al.} in which reds and colors that are more striking to the eye showed the greatest importance \cite{khosla2014image}. However our results both present blues and greens as less important. Another paper found that images with animate objects tend to be ranked as more funny than images with inanimate objects \cite{chandrasekaran2015humor}. Some of the colors found in the most popular memes may be colors that are more common in animate things like animals or human skin and hair tones. Black and off-white were also most present in the bottom 5\% of least popular memes, but other parts of the color profile differed. Greens and especially blues were more abundant in these memes, and some shades of orange and brown with large values in Figure \ref{fig:dank_colors} were not present at all in the least popular memes.

\begin{figure}[h!]
\centering
\includegraphics[width=\textwidth]{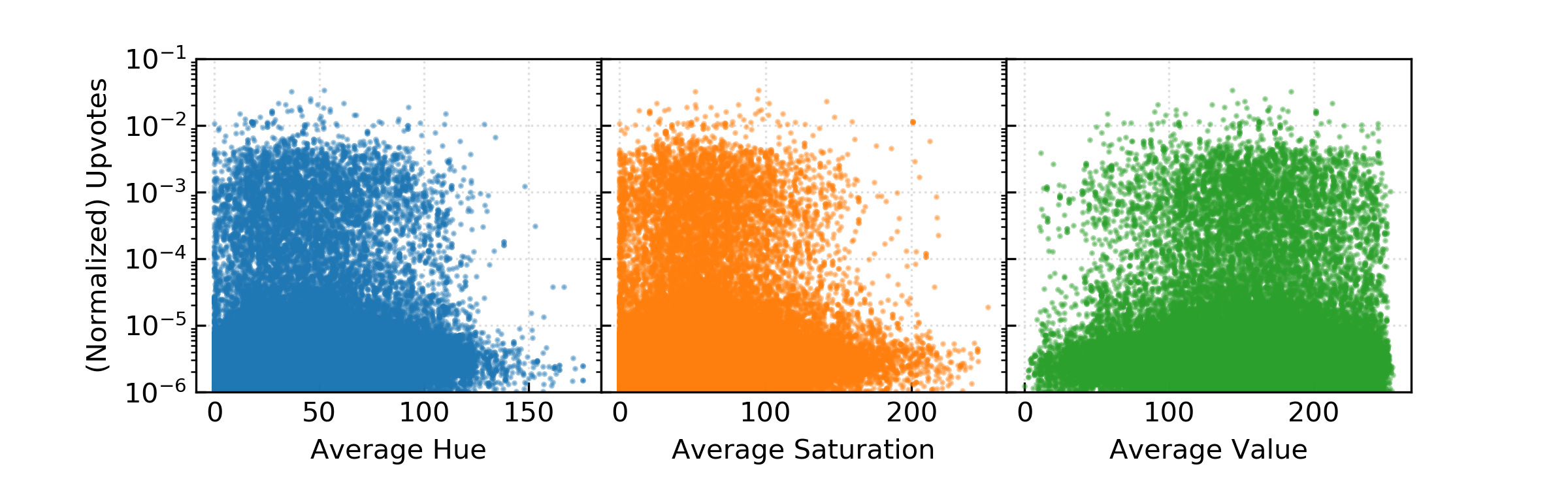}
\caption{\csentence{Average HSV and normalized upvotes.}
      The relationship between meme popularity, measured by normalized ups, and components of the HSV color space.}
\label{fig:hsv}
\end{figure}

In addition to colors, we extracted the average hue, saturation, and value components of the meme images. These are low level image features because while HSV mimics the way human (and now computer) vision works, these components of an image are not always obvious to the viewer. The relationship between these attributes and meme popularity is visualized in Figure \ref{fig:hsv}. Hue and saturation show a slight negative correlation with upvotes, indicating that yellow-green hued, less saturated images have a positive impact on popularity.  Value shows a slight positive correlation, indicating that images with higher value, more distinct and less dark colors may get more upvotes. These features tended to have significant predictive power in the machine learning models. 

\begin{figure}[h!]
\centering
\includegraphics[width=0.95\textwidth]{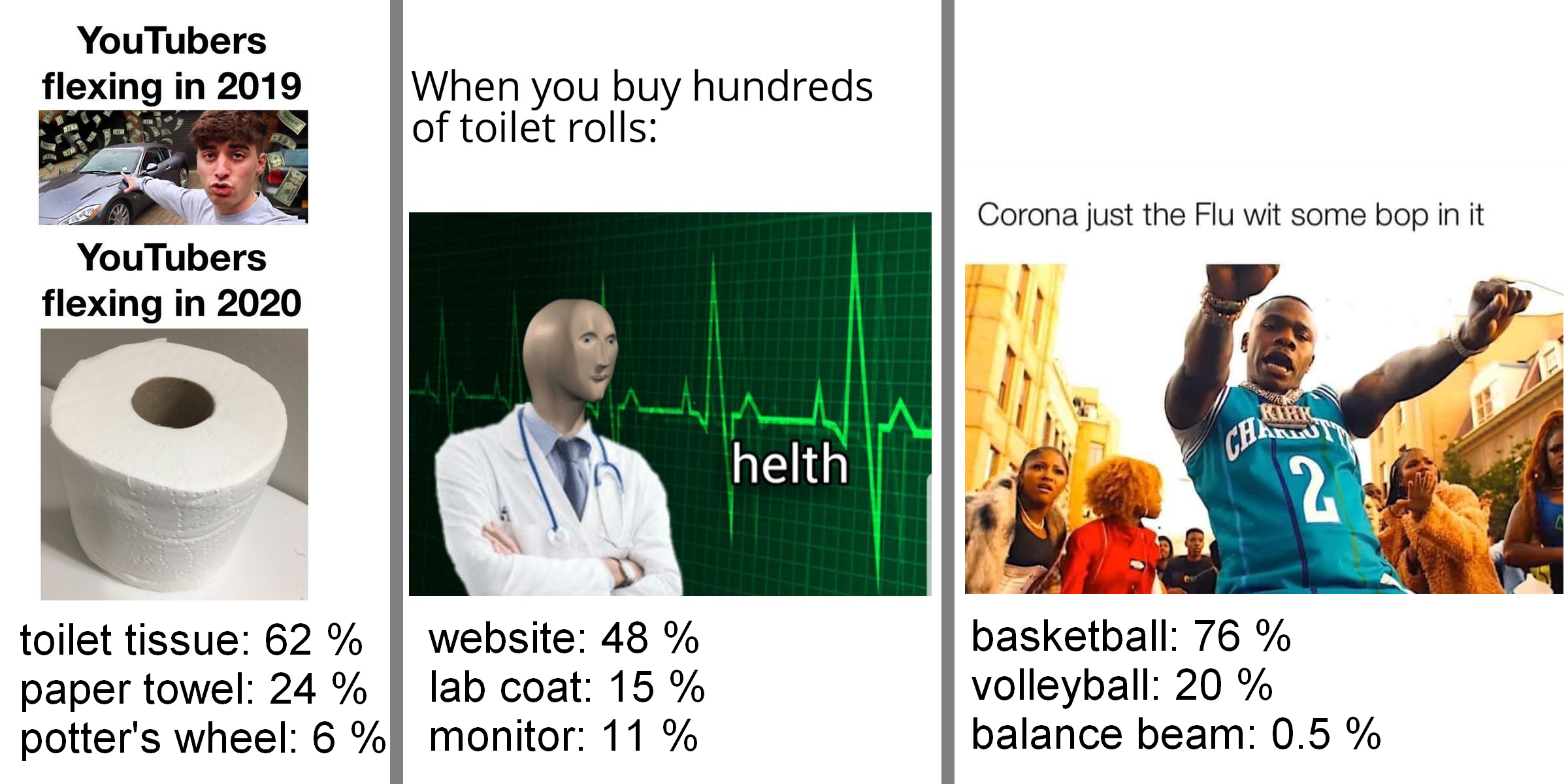}
\caption{\csentence{Neural Net High Level Image Features}
      We extracted high-level, categorical image features from the memes using the pre-trained keras neural network. Examples show the three most certain predicted image components and their probabilities by Keras VGG Neural Net. }
\label{fig:kerasmemes}
\end{figure}

We also analyzed high level image attributes that aim to describe the semantic meaning present in images. By processing the images with the pre-trained Keras' VGG-16 neural network, we were able to roughly identify what objects are present in the meme images \cite{chollet2015keras}. Figure \ref{fig:kerasmemes} shows the neural net's meme content predictions, with the associated probability of that prediction. This categorical data is not necessarily accurate, but does convey some level of information about the subject matter of each meme image. Table \ref{tab:VGG_table} lists which VGG-identified content was most common in the top 5 percent most popular memes and lower 5 percent least popular memes. These two columns list the top 10 unique values in each of these groups. The most and least popular memes also shared some VGG-identified content, such as the categories website, comic book, and book jacket. This is not surprising as many  memes are created using meme-making websites like Mimetic. The top ten shared content categories are listed in the third column of Table \ref{tab:VGG_table}. Many of the overlapping categories reflect the formatting of the meme and these were the most common categories identified by the VGG-16 neural network across all of the data records. Because terms about the image formatting were so common, we combined these terms into one category called \textit{formatted}. The neural net identified specific objects within the images less frequently, but these observations, as shown in Table~\ref{tab:VGG_table}, did tend to differ between the most and least popular 5 percent of memes.

While much of the VGG-identified content referred to miscellaneous items, some of the top categories related to the growing culture around the COVID-19 pandemic. Along with toilet\_tissue, lab\_coat and mask were within the top 40 most common VGG-identified components in the whole dataset. Many medical masks such as are worn to prevent the spread of COVID-19 were also misidentified as muzzels, gas masks or neck braces by the neural net. Thus these components were combined under one numerical attribute category, \textit{masks}.

\begin{table}[]
\begin{tabular}{|l|l|l|}
\hline
\textbf{Top 5\%} & \textbf{Bottom 5\%} & \textbf{Shared Content} \\ \hline
balance beam        & iron                  & web\_site               \\ \hline
military uniform    & academic gown         & comic\_book             \\ \hline
Chihuahua           & bow tie               & book\_jacket            \\ \hline
street sign         & matchstick            & envelope                \\ \hline
cash machine        & hair spray            & menu                    \\ \hline
crossword puzzle    & desktop computer      & scoreboard              \\ \hline
bakery              & digital clock         & laptop                  \\ \hline
refrigerator        & torch                 & toyshop                 \\ \hline
library             & barbell               & theater\_curtain      \\ \hline
toilet tissue       & basketball            & packet                \\ \hline
\end{tabular}
\caption{VGG-identified content in popular and unpopular memes}
\label{tab:VGG_table}
\end{table}

The categorical VGG-identified data was converted to numerical data in a number of ways. Upon observing that many of the VGG-identified objects belonged to similar categories, such as the meme formatting and masks mentioned before, we grouped these into 9 VGG content categories: \textit{animals}, \textit{formatted}, \textit{sports}, \textit{clothes}, \textit{masks}, \textit{technology}, \textit{violent content}, \textit{food and vehicles}. Note that some of the content identified in Image \ref{fig:kerasmemes} would be encoded in one or more of these categories. The categories were then one-hot-encoded into numerical features columns along with the next 8 most common VGG-identified content. These features were somewhat sparse, as the binary one-hot-encoding indicated whether or not a certain vgg prediction, or category, was found in the top three vgg content predictions for the meme. In addition to the binary features, we included the probabilities associated with the top 3 vgg content predictions for an additional 3 vgg related features. These probabilities tended to be ranked as important to the machine learning models discussed in the nextt section.

After these alterations to the raw image data, there were a total of 53 numerical image attributes. The abundance of features leaves room for fine-tuning and eliminating some of them to improve the models. Here, we suggested that certain colors and objects may be associated with viral memes, but the machine learning models will provide more clarity as to what characteristics are actually influential in determining the popularity of a meme. 

\subsection*{Gradient Boosting and Random Forest}
\label{boost-forest}

We selected Gradient Boosting and Random Forest models to perform the binary classification task of placing a memes in the \textit{dank} or \textit{not\_dank} categories. Both models are ensemble learners that benefit from the accumulated results of weak-learners. The models are trained and tested using the full array of data attributes listed in Table \ref{tab:features_table}, and discussed in the image and text analysis sections. They make predictions based on the same set of labels in which viral memes in the top 5\% of normalized upvotes are considered \textit{dank}, labeled 1, and the rest are \textit{not\_dank}, labeled 0. By observing how these ensemble models make their predictions we can garner insights about the most important features that make memes go viral. Using two models for this task will further validate our results.

Gradient boosting is an ensemble method of weak learners with the optimization of a loss function \cite{natekin2013}. Boosting models fit a new learner on the observations that the previous learner could not handle. The model serves as a good classifier for rank, which suited our binary classification task. The gradient boosting classifier of sklearn's ensemble package builds in a forward stage-wise manner, which means that a user-defined number of regression trees are fitted on the negative gradient of either the binomial or multi-nominal deviance loss function at each stage and the weighted sum of the learners will be the output \cite{pedregosa2011scikit}.

The Random Forest is an ensemble method made up of many decision trees. The success of the ensemble depends on the strength of the individual trees and the level of dependence between them. This model is a good choice for our data set because it performs well with a mix of categorical and continuous features, it can handle many features and large amounts of data without risk of over-fitting, and the tree structure is easily interpreted \cite{breiman2001randforest}. It is quite similar to the Gradient Boosting model, meaning they can be easily compared, and the differences between the models serve to reinforce our results, as our findings are replicated by two models.

\subsubsection*{Performance and Features Importance}

A limitation to the Random Forest and Gradient Boosting ensemble classifiers is that in their original form they do not perform that well with unbalanced data~\cite{Brownie2020imbalanced, liu2017addressing}. However, many methods for learning unbalanced data with these ensembles have been developed \cite{Chen2004UsingRF}. We modified the models to reduce the effect of skewed data, and generally improve the prediction results. Firstly, we used the BalancedRandomForestClassifier from imblearn as our Random Forest model. This classifier uses random undersampling to train on more balanced subsets of data by resampling data from the training set for each tree classifier in the ensemble. The distribution of positive and negative labels in the training sets can by controlled by the parameter \textit{sampling\_strategy} which represents the proportion of majority to minority class labels. Both the Random Forest and the Gradient Boosting models used 5-fold cross validation, the class weight parameter, and GridSearchCV from sklearn to fine-tune the classifiers' parameters.

\begin{figure}[h!]
\centering
\includegraphics[width=0.98\textwidth]{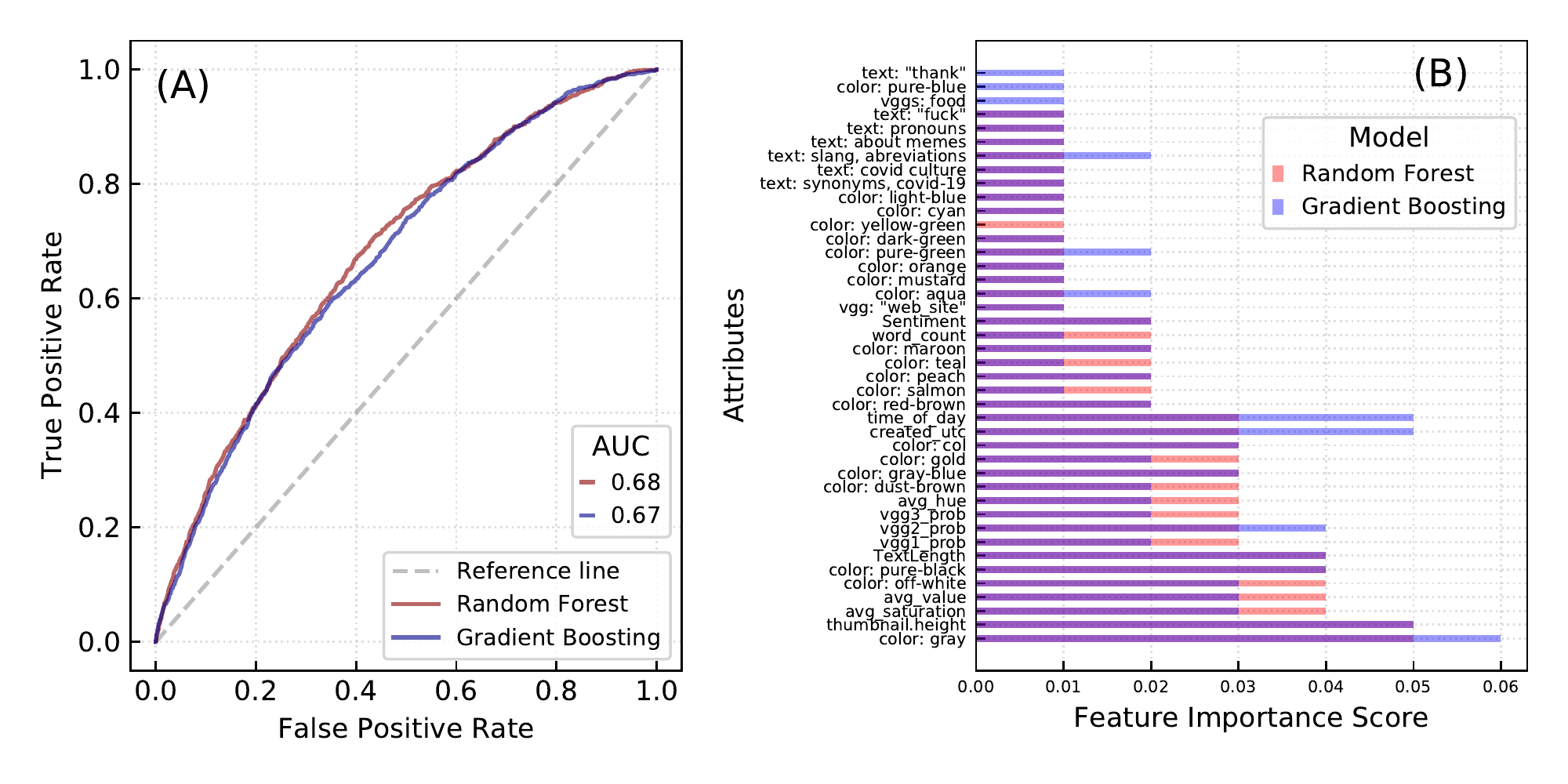}
\caption{\csentence{ROC Curve and Features Importance of Gradient Boosting and Random Forest models.} 
    A. The ROC curves of the models without undersampling techniques. B. Features importance for both models.}
\label{fig:rf_gb}
\end{figure}

Following these modifications, we split the data into a 53,843 record training set and 26,520 record test set, a 67\% - 33\% split. Both models predicted labels on the test set with an AUC of around 0.7 as shown in Figure \ref{fig:rf_gb}. Accuracy, recall, precision, and F-1 scores for the highest performing (Random Forest) model can been seen in Table \ref{tab:metrics_rf}, and scores for the gradient boosting model were quite similar. Both models performed poorly in precision. Consistently, the models predicted a larger proportion of positive labels than was realistic for the data set despite the measures we took. Some of the measures we took to to counteract this effect, such as re-assigning thresholds, were adjustable at a cost. Increasing the probability which was sufficient for a positive label would improve the model's precision but adjusting too much led the recall and accuracy scores to decrease.

The difficulty of predicting the imbalanced data, indicated by low precision scores, may be due to the lack of social network features. Perhaps while content-based features can predict whether a meme has a chance at going viral (has merit), social network features are what determines which of those memes actually do go viral. This supports Barab\'asi's theory on success in general in which merit is the first step to becoming successful, but social networks determine who among those with merit becomes a superstar \cite{barabasi2016success}.

\begin{table}[]
\begin{tabular}{|l|l|l|l|l|l|}
\hline
                    & Accuracy & Precision & Recall & F-1 Score & AUC    \\ \hline
Without Undersample & 0.6638   & 0.0854   & 0.5897 & 0.1492    & 0.6804 \\ \hline
Undersample train & 0.6043   & 0.0789   & 0.6486 & 0.1408    &  0.6689 \\ \hline
Undersample test + train    & 0.6366   & 0.6269    & 0.6742 & 0.6497    & 0.6814 \\ \hline
\end{tabular}
\caption{Metrics for the highest performing random forest models}
\label{tab:metrics_rf}
\end{table}

In addition to the modifications listen above, we tried a few undersampling methods. We performed a 67\% - 33\% train-test split for all models. The undersampling results for the best performing models, the Random Forest, are listed in Table \ref{tab:metrics_rf} and the results for the gradient boosting model were very similar. Using the random undersampler module from sklearn, we undersampled only the training data. This did not have a large effect on the models' performance, indicating that the other measures we took to counteract the imbalanced dataset, and generally improve results, were effective. We also tried undersampling both the test and train data, this improved the precision, and consequently F-1 scores immensely but had only a small effect on our AUC. (Of course, changing the distribution of the test set alters the nature of our prediction goals, therefore we do not report on the results of these tests extensively). We also note that while the precision value might look quite small without adjusting for the unbalanced distribution, but it means a more than 70\% improvement to random guessing dank memes.

The features importance plot in Figure \ref{fig:rf_gb} shows the relative importance of the data features from Table \ref{tab:features_table} for the Gradient Boosting and Random Forest models trained without undersampling. The two models showed similar features importance, with some variability. Additionally, many of the points explored in earlier sections about the features' relation to upvotes are reinforced by the importance scores. Simple features such as text length and image size (\textit{thumbnail.height}) showed great importance for predicting viral memes. The important colors in Figure \ref{fig:rf_gb} also align with the most abundant colors in Figure \ref{fig:dank_colors}. Gray, off-white, and pure-black are some of the most important colors for the model and are most abundant in viral memes. Figure \ref{fig:rf_gb} also indicates that overall more image-based than text-based features are important. However, this could be due to the fact that we included more image based than text based features overall, 53 as opposed to 38.

\subsubsection*{Incremental Predictive Power of Image and Text Features}

In addition to the most important features shown in Figure \ref{fig:rf_gb}, we investigated whether image or text features have more predictive power for determining viral memes. We used the Gradient Boosting and Random Forest models discussed previously with the full amount of train and test data. Differing from the earlier analysis, we trained four Gradient Boosting and four Random Forest models, each of the four with different subsets of features. The models are trained with image-only attributes, text-only attributes, both, and all attributes from Table \ref{tab:features_table} to show the incremental predictive benefit of these feature groups. The viral nature of memes makes predicting high performing memes more difficult, but since the skew is an inherent part of the data we decided against undersampling the data for this part of the analysis as changing the distribution alters the nature of our prediction question.

\begin{figure}[h!]
\centering
\includegraphics[width=\textwidth]{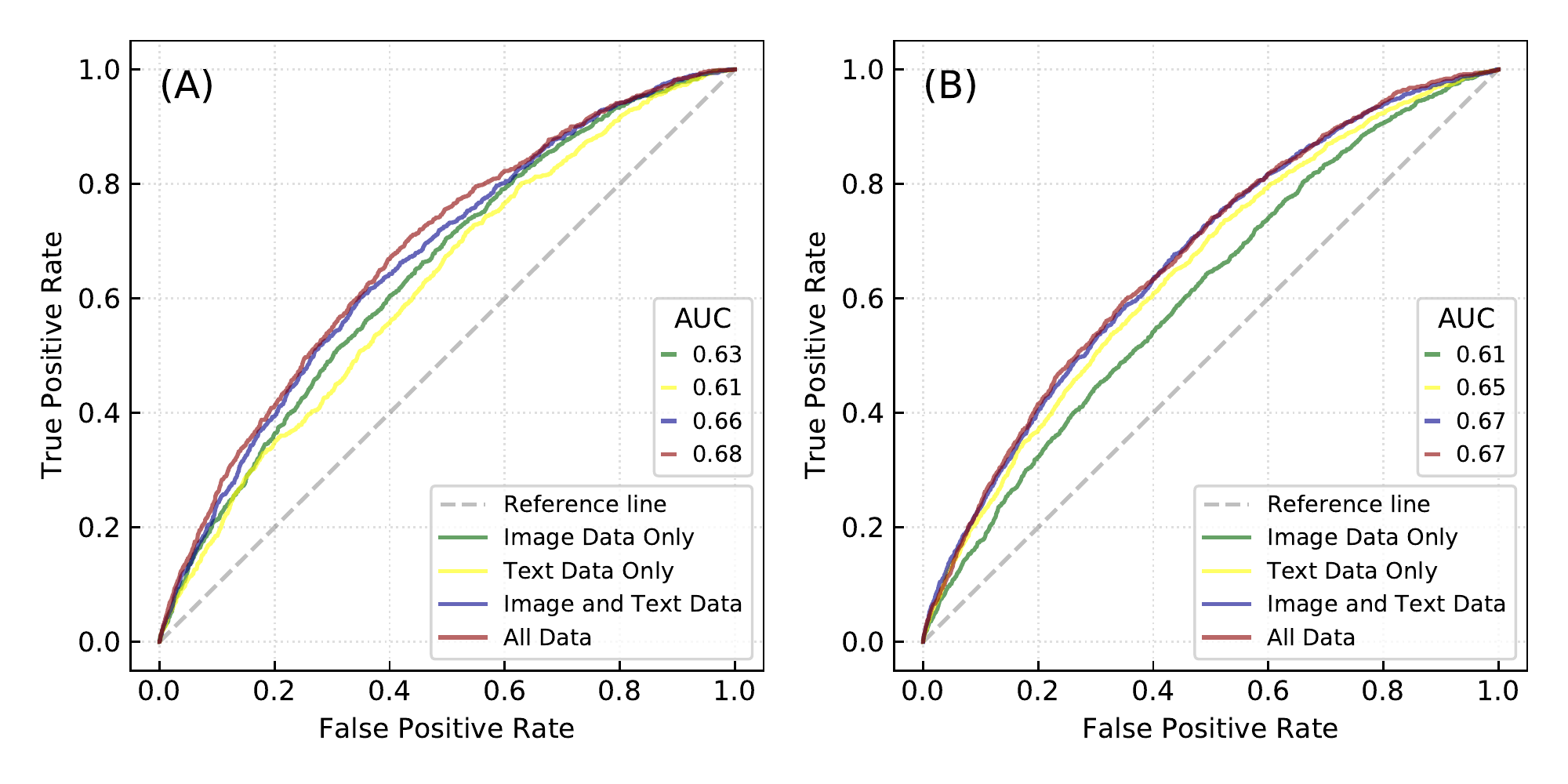}
\caption{\csentence{Incremental importance of image and text based features.}
      A. Random Forest model and B. Gradient Boosting model.}
\label{fig:ROCincremental}
\end{figure}

As for the previous models presented in Figure \ref{fig:rf_gb}, the models were trained with a set of 53,843 and test set of 26,520 data records, a 67\% - 33\% split. All of the modifications and fine-tuning, including class-weight and GridSearchCV, efforts used in those models are the same here. The exact feature description differs slightly between versions of these models due to fine-tuning efforts in which certain colors or processed words may have been eliminated if they showed no importance to the model. These slight differences did not disrupt the organization in which the four models had either only text related features, only image related features, both or all features including social network features scraped from the Reddit metadata such as subscribers.

Figure \ref{fig:ROCincremental} shows the results of the incremental predictive ability analysis. Not surprisingly, the model trained with all data outperformed the other models. This aligns with previous results in which text and network data held more predictive power for image popularity on Flickr \cite{khosla2014image}. It is impressive that adding only four network features (\textit{subscribers}, \textit{created\_utc}, \textit{is\_nsfw}, and \textit{time\_of\_day}) increased the AUC by 0.02 for the Random Forest Model. Given the results that social network features have shown in predicting meme popularity in past papers \cite{weng2012competition, weng2014net}, it is likely we would have seen a much greater increase in AUC if we had included other social network features, too. Surprisingly, it is not obvious whether image related or textual attributes have the stronger predictive power since the Random forest model performed better with the image related attributes, while the Gradient Boosting model performed better with textual attributes. However, it is clear that they both have incremental predictive power over each other in both models.

\subsection*{Transfer Learning with Convolutional Neural Network}
\label{transfer}

\subsubsection*{Convolutional Neural Network}
Convolutional Neural Network (CNN) is a class of artificial neural networks that has gathered attention in recent years due to its versatility and ability to achieve excellent performance in a multitude of problems. Among others, it had been used by computational linguistics to model sentences' semantics \cite{kalch2014sentence}, by radiologists to segment organs \cite{yamashita2018radio}, by ophthalmologists to identify diabetic retinopathy in patients~\cite{jama2016diabetic}. CNN's success lies in its architecture that allows it to learn inherent spatial hierarchies from its training data through recognizing and learning low-level patterns that build up to high-level patterns \cite{yamashita2018radio}. This ability to extract important features means that the CNN is able to identify different levels of image representation and capture the relevant ones in the training data, making this model family especially suitable for computer vision tasks \cite{jogin2018feature}. Past research has shown that CNN is also able to perform well when it comes to identify an image's popularity~\cite{khosla2014image}. Following this vein of research, in this section, we examine whether we could classify a meme's dankness solely based on raw image data, ignoring the attributes that we used in previous sections.

\subsubsection*{Sampling the Dataset}
The dataset contained approximately 76,000 downloadable images. Because of the imbalanced distribution in posts' upvotes as can be seen in Figure \ref{fig:normed_ups_distribution}, we chose to make the \textit{not\_dank} class to be the same size as the \textit{dank} class by randomly sampling from the 70,000+ images in the \textit{not\_dank} class. We then divided this sub-dataset into training set, validation set, and test set in the following ratio: 50\%, 25\%, 25\%. The exact number of images used in each set is shown in Table \ref{tab:subdataset_table}.

\begin{table}[H]
\begin{tabular}{|l|c|c|c|}
\hline
               & \textit{dank} & \textit{not\_dank} & Total \\ \hline
Training set   & 1,856                          & 1,856                               & 3,712 \\ \hline
Validation set & 928                            & 928                                 & 1,856 \\ \hline
Test set       & 929                            & 929                                 & 1,858 \\ \hline
Total          & 3,713                          & 3,713                               & 7,426 \\ \hline
\end{tabular}
\caption{Training, validation, and test set for the CNNs}
\label{tab:subdataset_table}
\end{table}

\subsubsection*{Transfer Learning}
Deep CNNs normally require a larger amount of training data than we had. Previous research has shown that in cases where there is limited training data, transfer learning is an effective method to significantly improve the performance of the neural network \cite{tamina2019transfer} as well as reduce overfitting \cite{han2018new}. Several transfer learning methods have been proposed throughout the years. Here, we adopt the method proposed by  Yosinski et al.~\cite{yosinski2014transfer}: using the top layers of the pre-trained CNNs as feature extractors, then fine-tuning the bottom layers with our own dataset, and adding a set of fully connected layers for prediction. The main reason we used this approach is the domain difference between our dataset and the ImageNet dataset which makes it necessary to retrain some of the last layers.

The pre-trained CNN models that serve as our feature extractors were trained using data from the ImageNet Large Scale Visual Recognition Challenge (ILSVRC)~\cite{olga2015imagenet}. This dataset consists of roughly 1.2 million training images, 50,000 validation images, and 150,000 testing images in 1,000 categories \cite{krizhevsky2017imagenet}. The pre-trained CNN models we picked -- namely, InceptionV3, VGG16, ResNet, Xception, MobileNet -- are top performers in previous ILSVRC competitions, and their weights trained with this dataset are all available in Keras \cite{kerasApp}.
Out of these models, VGG16, InceptionV3, and Xception proved to be the best performing feature extractors for our dataset. For further information about the models see Table \ref{tab:networksInfo}. We will provide more details about how we fine-tuned each of these neural networks in a later section.

\begin{table}[H]
\begin{tabular}{|l|c|c|c|}
\hline
Network name & Top 5 Accuracy on ImageNet & Parameters                                              & Depth \\ \hline
VGG16        & 0.901                      & 138,357,544 & 23    \\ \hline
Xception     & 0.945                      & 22,910,480  & 126   \\ \hline
InceptionV3 & 0.937                      & 23,851,784  & 159   \\ \hline
\end{tabular}
\caption{Information about each network}
\label{tab:networksInfo}
\end{table}

\subsubsection*{Image Data Augmentation}
Data augmentation is used to expand the dataset by generating and including similar yet slightly modified entries in the training process. In regard to image recognition tasks, the most traditional methods are to add noise or to apply affine transformations (e.g. translation, zoom, rotation, mirror, flip) \cite{suk2003affine}. Previous research has shown that this procedure could reduce error rate, helps with overfitting, and allows the model to converge faster \cite{david2001augmentation}. Yosinski et al. has reported that after augmenting the dataset with randomly translated images, their model see a decrease in error rate from 28\% to 20\% \cite{yosinski2015neural}. Another example is in the design of the VGG16 neural network that was among the winners of the ILSVRC 2014 competition, Simonyan and Zisserman also employed image augmentation techniques such as flipping the images, including randomly cropped patches of the images, or changing color intensity~\cite{simoyan2014VGG}. The authors claimed that this data augmentation helped decreased the error rate by 1\%. Similarly, our best three models are all trained using a dataset augmented with the following Keras ImageGenerator transformations:

\begin{itemize}
    \item Rescale the pixel values (between 0 and 255) to the [0, 1] interval.
    \item Zoom into the image randomly by a factor of 0.3.
    \item Rotate the image randomly by 50 degrees.
    \item Translate the image horizontally randomly by a ratio of 0.2 factor the image width.
    \item Translate the image vertically randomly by a ratio of 0.2 factor the image height.
    \item Shear the image randomly.
    \item Flip the image horizontally randomly.
\end{itemize}

\subsubsection*{Fine-tuning Strategies}
Since each network has different architectures, we needed to employ different fine-tuning strategies to each of them. The fine-tuning strategies we used are listed below:

\begin{itemize}
    \item For VGG16, freezing the first three convolution blocks, fine-tuning the weights of all the other layers (in the two other convolution blocks in the network, plus the last three fully-connected layers).
    \item For Xception, freezing the weights of all convolutional layers, and fine-tuning the weights of only the last three fully-connected layers.
    \item For InceptionV3, freezing the weights of all the layers up until the ``mixed7" layer, then fine-tuning the rest of the layers in the InceptionV3 network plus the last three manually-defined fully-connected layers.
    \item For all networks, dropout is implemented after the first and the second layer of the last three fully-connected layers. 
    \item For all networks (VGG16, Xception, InceptionV3), the last ``softmax" layer is removed, and replaced by a ``sigmoid" layer for prediction. 
    \item The images are resized to fit the default input size for each of the network (299x299 for Xception and InceptionV3, 224x224 for VGG16).
    \item All networks include ReduceLROnPlateau functionality from Keras that reduces the current learning rate by 25\% whenever the validation accuracy does not increase in the span of 3 epochs. 
\end{itemize}

\subsubsection*{Results}
\begin{figure}[h!]
\centering
\includegraphics[width=\textwidth]{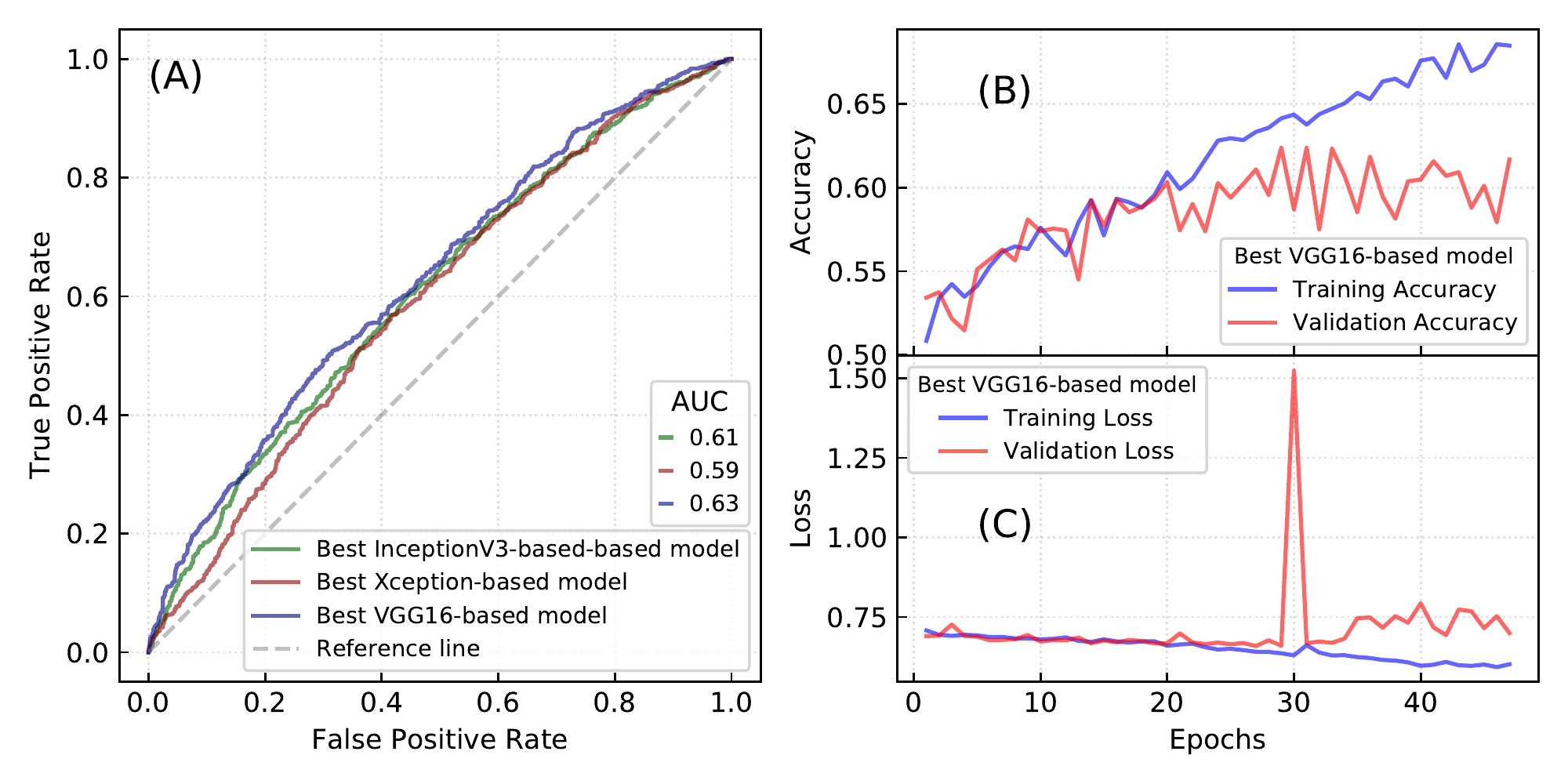}
\caption{\csentence{ROC curves of the 3 best CNN models and the training curves of the best model.} ROC curves and AUC scores (A) of the best models based on pre-trained CNN models. The accuracy (B) and loss (C) during training of the best VGG16-based model.}
\label{fig:roc_3_best_cnn}
\end{figure}

All of the neural networks tested for this research were evaluated on the test set using several metrics (Accuracy, Precision, Recall, and F-1 Score). The results are recorded in Table \ref{tab:neuralResults}. We also calculated the ROC curve of the best 3 models along with their AUC scores which are shown in Figure \ref{fig:roc_3_best_cnn}.A. Figure \ref{fig:roc_3_best_cnn}.B and C show the change in the training and validation accuracy and loss during fine-tuning the VGG16-based model, which produced the best AUC score.

\begin{table}[h]
\begin{tabular}{|l|c|c|c|}
\hline
\multicolumn{1}{|c|}{Metrics} & VGG16  & Xception & InceptionV3 \\ \hline
Accuracy                      & 0.5721 & 0.5608   & 0.5327      \\ \hline
Precision                     & 0.5732 & 0.5752   & 0.5362      \\ \hline
Recall                        & 0.5721 & 0.5608   & 0.5327      \\ \hline
F-1 Score                     & 0.5706 & 0.5388   & 0.5212      \\ \hline
\end{tabular}
\caption{Test set performance for each neural network}
\label{tab:neuralResults}
\end{table}

\noindent
From Table \ref{tab:neuralResults} and Figure \ref{fig:roc_3_best_cnn}, we can conclude that the VGG16-based model seems to slightly outperform the other models, while the Xception-based model comes in second, and the InceptionV3-based model in third place. We can also conclude that the best neural network (AUC=0.63) performs equally with the best performing ensemlbe model trained on hand-crafted image features (AUC=0.63).



\begin{figure}[h!]
\centering
\includegraphics[width=0.98\textwidth]{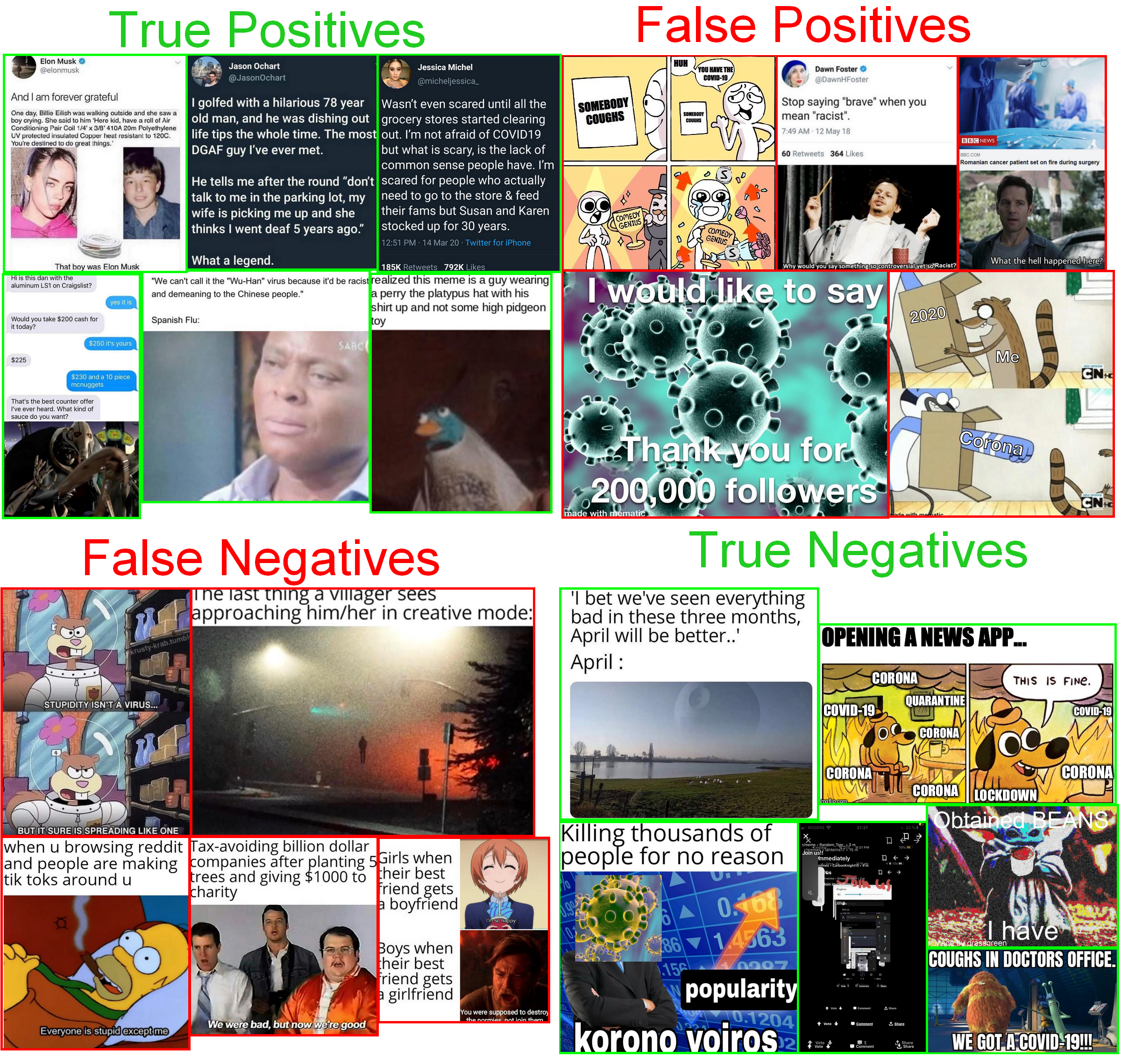}
\caption{\csentence{A confusion matrix with example memes according to the CNN model.} The green bordered parts show True Positive and True Negative instances where the model prediction is the same as the real target label. The red-bordered parts are False Positive and False Negative instances where the predicted and true class labels do not match.}
\label{fig:cnn_confusion_matrix}
\end{figure}

From our experiments with different models, we have observed that using image augmentation helps with making the models converge faster and achieve a higher accuracy. Fine-tuning the last few layers of the CNN models with the transfer learning methodology also improved the performance of our models. The over-fitting issues we encountered were depressed by adding a dropout rate between layers and reducing the learning rate between epochs. Although we have experimented with several models and parameters, the model performances show that it is hard to predict the \textit{dankness} of image-and-text memes using the image content alone. This finding is in line with similar previous research where image-content has smaller significant compared to social contexts and other features for predicting image popularity on Flickr \cite{khosla2014image}. The difficulty of content based popularity prediction of memes is also illustrated by the memes where the model prediction does not match the true class label (see Figure \ref{fig:cnn_confusion_matrix}). It is a difficult task to tell the true class of these memes both for humans and machines.

\section*{Conclusion}

In this paper, we analyzed image-with-text memes collected from Reddit. Using machine learning models we investigated whether viral memes can be predicted based on their content alone. We considered the problem as a binary classification task defining viral memes as the top 5\% of all posts in terms of upvotes. Our best performing model is a random forest model that performs moderately well with an AUC of 0.6804, accuracy of 0.6638, precision of 0.0854, recall of 0.5897. While the precision value might seem quite low at first sight, it is a 70\% improvement to random guessing dank memes.

Moreover, we studied the most important features and we found that gray content, image size, saturation, and text length have the greatest impact on the prediction. While there was a lot of COVID-19 related content in the dataset overall, visible in the word cloud, and some vgg-identified image content, features related to COVID-19 proved less important to the performance of the models. Thus, we estimate that while memes often reflect pressing world issues, the presence of this sort of content has little impact on whether memes will go viral. We also investigated the predictive and incremental predictive power of image and text features. While we cannot conclude whether image related or textual attributes are the stronger predictors of a meme's success, we have shown that they both have incremental predictive power over each other. If we use only the images as an input with a convolutional neural network we can reach AUC=0.63, and that agrees with the performance of the best performing random forest model trained on hand-crafted image features. Comparing our results with other works where social network and community features were also used for predicting popularity \cite{weng2012competition, weng2014net}, we can conclude that while the content-based analysis can also predict success with reasonable efficiency, social network features could improve the performance significantly. While content based features could predict memes with merit, social network features determine which among those with merit actually go viral.

It is also fair to acknowledge some limitations of this study. Due to the the short time period in which we collected data - in the intense moment at the beginning of the coronavirus outbreak - our results cannot necessarily be universally generalized. However, we believe that many of our findings are relevant for meme popularity in general. Moreover, the short time period of the collected data did not allow us to study the temporal and dynamic aspects of meme success or identify  so-called "sleeping beauties". The latter is a phenomenon of information spread in which a meme will remain unnoticed for a long period and then suddenly spike in popularity long after it was originally posted \cite{zhang2016sleepingbeauty}. We propose these aspects of meme popularity prediction for future research. Furthermore, an other stream of relevant future research would be to analyze memes inspired by COVID-19 alone.

\section*{List of abbreviations}
\begin{itemize}
    \item HSV: hue saturation, value
    \item OCR: Optical Character Recognition
    \item LSTM: Long Short Term Memory
    \item ROC: receiver operating characteristic
    \item AUC: area under the curve
    \item CNN: Convolutional Neural Network
    
\end{itemize}


\begin{backmatter}

\section*{Availability of data and materials}

The datasets analysed during the current study are available in the GitHub repository: \url{ https://github.com/dimaTrinh/dank_data}

\section*{Competing interests}
  The authors declare that they have no competing interests.

\section*{Author's contributions}
    MDT have conceived the study. KB and RM reviewed the literature. TR and MDT collected the data. KB and EL performed the text analysis. KB performed the image analysis. KB and TR trained and analyzed the Gradient Boosting and Random Forest models. MDT trained and analyzed the Convolutional Neural Network Model. NB designed the figures and helped supervise the project. RM supervised the project. All authors contributed to the writing of the manuscript, read and approved the final version.

\section*{Acknowledgements}
Not applicable
  
\section*{Funding}
The research reported in this paper and carried out at the BME has been supported by the NRDI Fund based on the charter of bolster issued by the NRDI Office under the auspices of the Ministry for Innovation and Technology. The research of RM was also supported by the NKFIH K123782 research grant.
  

\bibliographystyle{bmc-mathphys} 
\bibliography{bmc_article}      







\end{backmatter}
\end{document}